\documentclass{jltp}

\usepackage{graphicx}
\usepackage{amsmath,amsthm,amssymb}

\title{Imaginary Squashing Mode Spectroscopy of Helium Three B}

\author{J.P. Davis, H. Choi, J. Pollanen, and W.P. Halperin}

\address{Department of Physics and Astronomy,\\Northwestern University, Evanston, Illinois 60208, USA}

\runninghead{J.P. Davis \emph{et al.}}{Imaginary Squashing Mode
Spectroscopy of Helium Three B}

\begin{document}

\maketitle

\begin{abstract}
We have made precision measurements of the frequency of a
collective mode of the superfluid $^{3}$He-B order parameter, the
$J=2^{-}$ imaginary squashing mode. Measurements were performed at
multiple pressures using interference of transverse sound in an
acoustic cavity. Transverse waves propagate in the vicinity of
this order parameter mode owing to off-resonant coupling.  At the
crossing of the sound mode and the order parameter mode, the sound
wave is strongly attenuated. We use both velocity and attenuation
measurements to determine precise values of the mode frequency
with a resolution between 0.1\% and 0.25\%.

\end{abstract}

\section{INTRODUCTION}
    Of the $J=2$ collective modes in $^3$He-B, the
imaginary squashing (ISQ) mode has been measured with considerably
less precision than the real squashing (RSQ) mode. The ISQ mode
couples strongly to longitudinal sound and therefore has been
difficult to resolve\cite{Hal90}. As a result, detailed analysis
of the corrections to the collective mode spectrum of the type
performed on RSQ mode data has not been possible for the ISQ mode.
 To first order, these modes have frequencies proportional to the
gap:
\begin{equation}\label{freq}
    \Omega_{2^{\pm}} = a_{\pm}\Delta^{+}(T,P),
\end{equation}
where $a_{+} = \sqrt{8/5}$ (RSQ) and $a_{-} = \sqrt{12/5}$ (ISQ).
Measurements have shown that the RSQ mode frequencies deviate from
this value and Sauls and Serene\cite{Sau81} have shown that this
is because the collective mode frequencies are strongly influenced
by both quasi-particle interaction effects and $f$-wave pairing.
We have made measurements using the interference of transverse
sound in an acoustic cavity to resolve the ISQ mode frequencies
with greater precision than previously reported.
\begin{figure}[t]
\centerline{\includegraphics[width=4in]{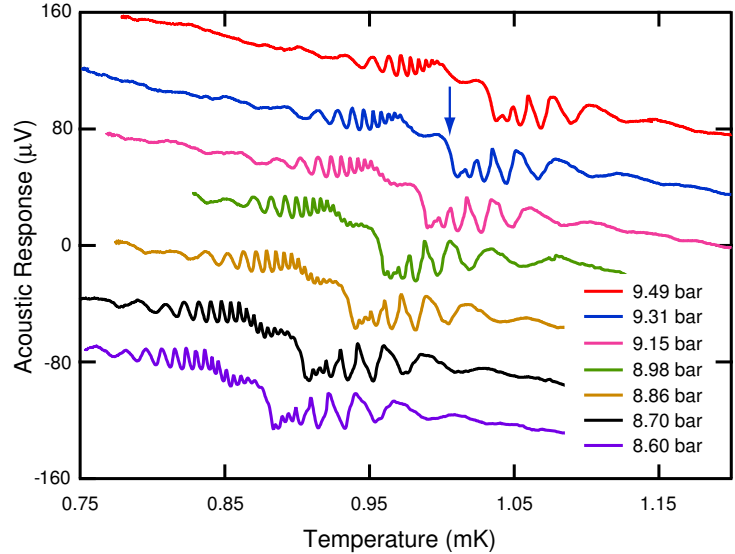}}
\caption {\label{data}(Color on-line) Interference of transverse
sound near the imaginary squashing mode using 99.9484 MHz sound at
pressures near 9 bar.  The cessation of oscillations and bend in
the acoustic impedance marks the location of the imaginary
squashing mode. As the pressure is changed the location of the ISQ
mode changes correspondingly. Detailed analysis of the
oscillations, Fig.~\ref{period}, allows a precise identification
of the mode crossing, indicated by the arrow for 9.31 bar.}
\end{figure}
\noindent
\section{RESULTS AND DISCUSSION}
An acoustic cavity is formed from an \emph{AC}-cut quartz
transducer and a reflecting plate separated by monodispersed latex
microspheres. Using an electrical-impedance spectrometer in a
bridge configuration, we have measured the transverse acoustic
impedance of superfluid $^3$He as a function of temperature at
high frequencies. The acoustic impedance is given by\cite{Hal90}
$Z=\rho / \omega q=\rho C$, where $C$ is the complex phase
velocity and $q = \omega / c + i\alpha$. The acoustic impedance is
simultaneously sensitive to changes in the (normal fluid) density,
$\rho$; phase velocity, $c$; and attenuation, $\alpha$.
Fig.~\ref{data} displays the acoustic impedance as the temperature
is lowered through the imaginary squashing mode at a fixed
transverse sound frequency. As the temperature nears that of the
ISQ mode, transverse sound begins to propagate due to off-resonant
coupling to the mode\cite{Moo93,Kal93,Lee99}, as described in
Eq.~(\ref{dispersion}). This results in the interference pattern
seen on the high temperature side of the mode.  The attenuation
decreases with decreasing temperature and the interference
oscillations become larger.  With further decrease in temperature
the pattern is quenched owing to resonant coupling to the ISQ,
which increases the attenuation.  The standing transverse waves
reappear on the low temperature side of the ISQ mode and finally
die off as the off-resonant coupling becomes weak. This is the
first time that transverse sound has been detected at temperatures
below the ISQ.
\begin{figure}[t]
\centerline{\includegraphics[width=2.5in]{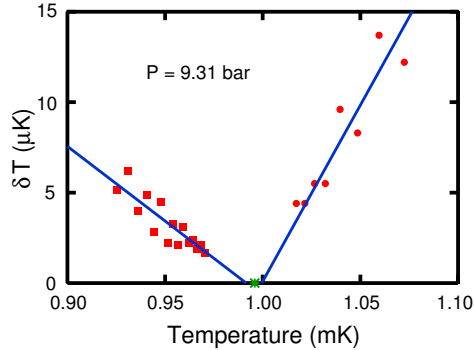}}
\caption {\label{period}(Color on-line) The period shift analysis
for determining the temperature of the ISQ mode crossing at
99.9484 MHz and 9.31 bar. High and low temperature data are fit
independently to give the temperatures of zero period shift, which
are then averaged to assign the ISQ mode crossing.  This
temperature is indicated in Fig.~\ref{data} by the arrow.}
\end{figure}
\noindent

\begin{figure}[t]
\centerline{\includegraphics[width=4in]{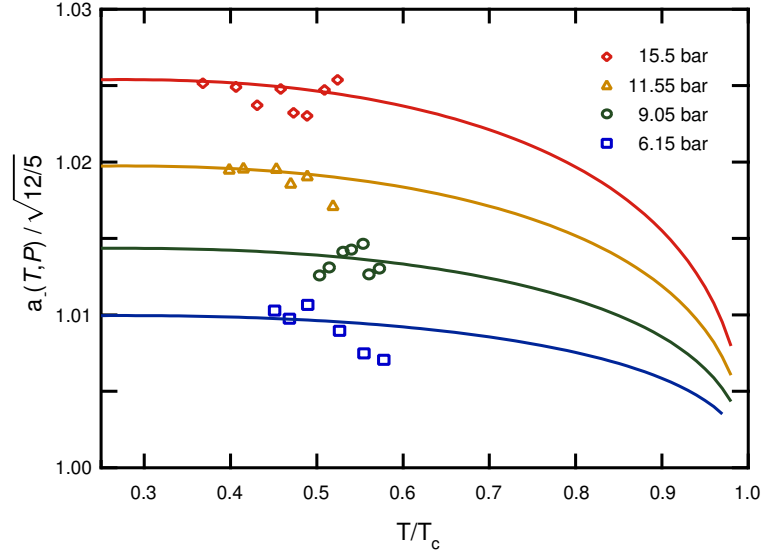}}
\caption {\label{results}(Color on-line) Imaginary squashing mode
crossings divided by $\sqrt{12/5}$ as determined by the period
shift analysis at four normalized pressures.  The curves are fit
to the data using the theory of Sauls and Serene\cite{Sau81}.}
\end{figure}

The proximity to the ISQ mode determines the phase velocity,
resulting in the period of the oscillations becoming smaller upon
approaching the mode crossing. We have used the period of the
oscillations, to precisely determine the crossing of the
transverse sound mode with the imaginary squashing mode.  We plot
the temperature difference between the peak and trough of the
acoustic impedance versus their average temperature for one
particular frequency and pressure in Fig.~\ref{period}. The data
on either side of the mode are independently fit to determine at
what temperature the period goes to zero. The position of the ISQ
mode at the measurement frequency is assigned to be the average of
these two temperatures. This is repeated for all pressures and
frequencies. In Fig.~\ref{results}, we plot the ISQ mode as
$a_{-}(T,P)/\sqrt{12/5}$ versus reduced temperature. Strong
coupling is taken into account, by using the weak-coupling plus
form\cite{Rai76} of the gap, $\Delta^{+}(T,P)$, as tabulated in
Ref. \onlinecite{Hal90}.  To take data at a fixed acoustic
frequency we can tune the gap amplitude by varying the pressure
over a small range.  For the lowest frequency data the pressure
variation was $\pm$ 0.5 bar, increasing to $\pm$ 1 bar at the
highest frequency.  Data taken in this way can be converted to a
constant pressure, using Eq.~(\ref{freq}). The data in
Fig.~\ref{results} is plotted for fixed pressures. The temperature
of the mode crossings have an uncertainty between 5 and 15 $\mu
K$, yielding an uncertainty in $a_{-}(T,P)/\sqrt{12/5}$ between
0.1\% and 0.25\%.
\noindent

In order to confirm that our determinations of the ISQ mode
crossings are accurate, we compare our data with the theory of
Moores and Sauls\cite{Moo93}. They have shown that the dispersion
of transverse sound is given by,
\begin{multline}\label{dispersion}
    \left(\frac{\omega}{q v_f}\right)^{2} = \frac{F_1^{s}}{15}[1-\lambda(\omega,T)] + \frac{2F_1^{s}}{75}\lambda(\omega,T)\\
    \times\frac{\omega^{2}}{(\omega+i\Gamma(T))^{2} - \Omega_{2^{-}}(T,P)^{2} - \frac{2}{5}q^{2}v_f^{2}},
\end{multline}
where $v_{f}$ is the Fermi velocity, $\omega$ is the measurement
frequency, $\Omega_{2^{-}}(T,P)= a_{-}(T,P)\Delta^{+}(T,P)$ is the
ISQ mode frequency, $\lambda(\omega,T)$ is the Tsuneto function
and $F_1^{s}$ is a Fermi liquid parameter derived from measurement
of the heat capacity. $\Gamma(T)$ is the width of the mode, with
an approximate form\cite{Moo93} of
$\Gamma(T)\simeq\Gamma_{c}\sqrt{T /
T_{c}}e^{-\frac{\Delta(T)}{T}}$ and $\Gamma_{c}\sim10^{6}-10^{7}$
Hz.  It is this mode width that allows for transverse sound
propagation at temperatures below the ISQ mode.  Solving
Eq.~(\ref{dispersion}) at temperatures low compared with $T_c$,
where the first term goes to zero, one finds that there are no
corrections to the mode frequencies because of non-zero $q$,
so-called dispersion corrections. The next correction that remains
is of order $\Gamma(T)^{2} / \omega$ which is considerably smaller
than the uncertainty in our measurement.
\begin{figure}[t]
\centerline{\includegraphics[width=3.125in]{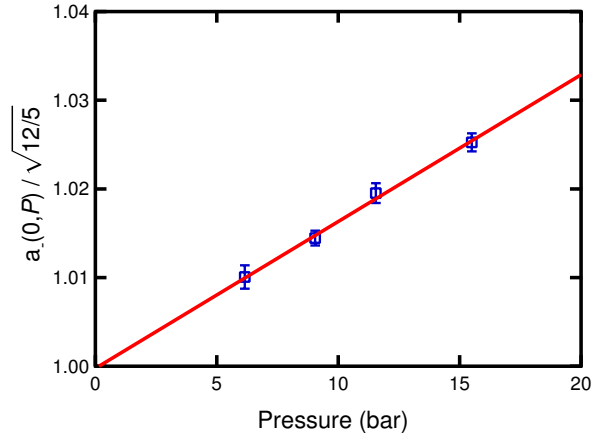}}
\caption {\label{a}(Color on-line) Pressure dependance of
$a_{-}(0,P)/\sqrt{12/5}$, the zero temperature intercept of the
fits to the imaginary squashing mode frequencies shown in
Fig.~\ref{results}. The error bars reflect the rms deviation of
the data points from the curves in Fig.~\ref{results}. The line
through the points is an unconstrained fit to the data and shows
that at zero pressure, $a_{-}(0,P)/\sqrt{12/5}= 0.9998 \pm
0.0018$.}
\end{figure}
\noindent

Note that the ISQ mode frequencies in Fig.~\ref{results} lie above
that given by Eq.~(\ref{freq}). This is the opposite of the RSQ
mode\cite{Hal90}, where the mode frequencies lie below the value
of $a_{+}(T,P)=\sqrt{8/5}$.  Sauls and Serene\cite{Sau81} have
shown that normal state quasi-particle interactions as well as
pairing interactions higher than $p$-wave, namely $f$-wave,
influence the collective mode frequencies. For the sake of
brevity, these higher order effects will be discussed in a future
paper. In Fig.~\ref{a}, we plot the zero temperature intercept of
these fits $a_{-}(0,P)/\sqrt{12/5}$, as a function of pressure.
The line is an unconstrained fit to the data and demonstrates the
internal consistency of the analysis.  Note that at zero pressure
$a_{-}(0,0)/\sqrt{12/5}= 0.9998 \pm 0.0018$. One expects that
$a_{-}(0,0)/\sqrt{12/5}=1$ if quasiparticle and $f$-wave
interactions are negligible and corrections to the temperature
scale\cite{Gre86} are insignificant.  Use of the BCS gap results
in $a_{-}(0,0)/\sqrt{12/5}= 1.014$.

In summary, we have made new precision measurements of the
imaginary squashing mode as a function of pressure.  This
precision derives from our use of the interference of high
frequency transverse sound in an acoustic cavity and comparison
with the theory of Moores and Sauls\cite{Moo93}. The measurements
reveal that the mode frequencies lie above the value of
$\sqrt{12/5}\Delta^{+}(T,P)$. Extrapolating fits of the zero
temperature ISQ mode frequencies to zero pressure recovers this
value. In a future paper, we will discuss the implications of
these measurements for the $f$-wave pairing interaction strength.

\section*{ACKNOWLEDGMENTS}
 We acknowledge support from the National
Science Foundation, DMR-0244099 and would also like to thank J.A.
Sauls for helpful discussions.


\begin{thebibliography}{99}


\bibitem{Hal90}
W.P. Halperin and E. Varoquaux, in \emph{Helium Three}, ed. by
W.P. Halperin and L.P. Pitaevskii (Elsevier Science Publishers,
Amsterdam 1990).

\bibitem{Sau81} J.A. Sauls and J.W. Serene, Phys.
Rev. B \textbf{23}, 4798 (1981).

\bibitem{Moo93}
G.F. Moores and J.A. Sauls, Jour. of Low Temp. Phys \textbf{91},
13 (1993).

\bibitem{Kal93}
S. Kalbfeld, D.M. Kucera, and J.B. Ketterson, Phys. Rev. Lett.
\textbf{71}, 2264 (1993).

\bibitem{Lee99}
Y. Lee, T.M. Haard, W.P. Halperin and J.A. Sauls, Nature
\textbf{400}, 431 (1999).

\bibitem{Rai76}
D. Rainer and J.W. Serene, Phys. Rev. B \textbf{13}, 4745 (1976).

\bibitem{Gre86}
D.S. Greywall, Phys. Rev. B \textbf{33}, 7520 (1986).
\end{thebibliography}
\end{document}